\documentclass[aps,prb,superscriptaddress,twocolumn,floatfix]{revtex4-2}
\usepackage[utf8]{inputenc}
\newcommand{\be}{\begin{equation}}
\newcommand{\ee}{\end{equation}}
\newcommand{\bea}{\begin{eqnarray}}
\newcommand{\eea}{\end{eqnarray}}
\usepackage{amsmath}
\usepackage{amssymb}
\usepackage{amstext}
\usepackage{mathrsfs}
\usepackage{graphicx}
\usepackage{hyperref}
\usepackage{makecell}
\usepackage{simpler-wick}
\usepackage{float}
\usepackage{xcolor}
\usepackage{ulem}
\usepackage{comment}
\usepackage{mathtools}
\usepackage[caption=false]{subfig}
\usepackage{dsfont}
\hypersetup{
  colorlinks   = true, 
  urlcolor     = blue, 
  linkcolor    = blue, 
  citecolor   = red 
}

\begin{document}

\title{On the Adequacy of the Dynamical Mean Field Theory for Low Density and Dirac Materials}
\author{Anqi Mu}
\affiliation{Department of Physics, Columbia University, New York, NY 10027, USA}
\author{Zhiyuan Sun}
\affiliation{Department of Physics, Harvard University, Cambridge, MA 02138, USA}
\affiliation{State Key Laboratory of Low-Dimensional Quantum Physics and Department of Physics, Tsinghua University, Beijing 100084, P. R. China}
\author{Andrew J. Millis}
\affiliation{Department of Physics, Columbia University, New York, NY 10027, USA}
\affiliation{Center for Computational Quantum Physics, Flatiron Institute, 162 5th Avenue, New York, NY 10010, USA}

\date{\today}

\begin{abstract}

The qualitative reliability of the dynamical mean field theory (DMFT) is investigated for systems in which either the actual carrier density or the effective carrier density is low, by comparing the exact perturbative and dynamical mean field  expressions of electron scattering rates and optical conductivities. We study two interacting systems: tight binding models in which the chemical potential is near a band edge and Dirac systems in which the chemical potential is near the Dirac point. In both systems it is found that DMFT underestimates the low frequency, near-Fermi surface  single particle scattering rate by a factor proportional to the particle density. The quasiparticle effective mass is qualitatively incorrect for the low density tight binding model but not necessarily for Dirac systems.  The dissipative part of the optical conductivity is more subtle:  in the exact calculation vertex corrections, typically neglected in DMFT calculations, suppress the low frequency optical absorption, compensating for some of the DMFT underestimate of the scattering rate.  The role of vertex corrections in calculating the conductivity for Dirac systems is clarified and a systematic discussion is given of the approach to the Galilean/Lorentz invariant low density limit. Relevance to recent calculations related to Weyl metals is discussed.

\end{abstract}

\maketitle

\section{Introduction}
\label{sec:intro}
The dynamical mean field approximation  is an extraordinarily useful tool for computing and understanding the properties of correlated electron systems in both model system \cite{Georges96} and real materials \cite{Kotliar06,Held07} contexts.  The approximation becomes exact in an infinite dimensional or infinite coordination number limit in which the electron self energy is momentum independent and it is believed to be reasonably accurate when the momentum dependence of the self energy is weak relative to the frequency dependence, as occurs for example in lattice models with typical band fillings. However, for low electron density one may expect the physics to be described by a theory with an effective Galilean or Lorentz invariance, implying significant momentum dependence of the electronic self energy and therefore a breakdown in the reliability of the DMFT approximation. While it has been generally understood since the earliest days of the dynamical mean field approximation that the approximation was likely to break down near band edges \cite{Georges96} a systematic study has been lacking.

`Dirac' and `Weyl' materials, semimetals in which a protected band crossing creates Dirac points at which the  density of states vanishes, are of intense recent interest \cite{liang2015ultrahigh}. These materials typically have a high density of electrons (roughly one per unit cell in the conduction bands) but are in a sense low density systems because if the fermi level is near the Dirac point and there are no extended Fermi surface regions, the density of mobile electrons may be small. Dynamical mean field theory has nevertheless been applied \cite{Wang17,Wagner21,Poli23} and it is therefore of interest to understand the limits of applicability of the dynamical mean field approximation in this case.

In a previous paper \cite{Mu22} we compared DMFT and perturbatively exact expressions for the optical conductivity of the two dimensional Hubbard model over a wide density range. Here we extend the analysis to three dimensional Hubbard and Dirac systems, consider the self energy in more detail and provide a systematic examination of the low density limit. We find that the dynamical mean field approximation qualitatively underestimates the low frequency part of the imaginary part of the self energy by a factor proportional to the particle density, calling into question some of the results obtained in references \cite{Wagner21,Poli23}. For the two and three dimensional tight binding models (throughout the paper tight binding model refers to one band Hubbard model) with chemical potential near the band edge, we find that DMFT also fails to capture the quaisparticle mass enhancement correctly.  However, for  three dimensional Dirac systems, dynamical mean field theory may provide a reasonable approximation to the real part of the electron self energy and therefore to the quasiparticle mass enhancement and  the critical interaction strength required to drive a Mott transition. 

The important features of our result can be understood from a scaling argument. In a system with  a continuous translation invariance relevant energy scales are the chemical potential, the frequency and the temperature. At temperature $T=0$ we expect
\begin{equation}
\mathrm{Im}\Sigma(\omega,k)=\mu CI\left(\frac{\omega}{\mu},\frac{k}{k_F}\right),
\label{eq:scaling}
\end{equation}
with $C$ a constant determined by the interactions and density of states, $I$ a scaling function and Fermi momentum  $k_F\propto \mu^\lambda$ ($\lambda=\frac{1}{2}$ for Galilean and $=1$ for Lorentz-invariant systems). A continuous translation invariance can emerge as a low density limit of a lattice model, in which case one would expect Eq.~(\ref{eq:scaling}) to hold in a low density/low frequency limit/small k limit ($\mu\rightarrow 0$, $\omega/\mu$ and $k/k_F$ of order unity).  
In a translation invariant Fermi liquid one expects that at $\omega\rightarrow 0$ for kinematic reasons the leading $\omega^2$ term in $\mathrm{Im} \Sigma$ vanishes for $k>3k_F$, so at low frequency $\mathrm{Im}\Sigma$ has a strong momentum dependence on the scale of $k_F$. The dynamical mean field approximation in effect approximates the full frequency and momentum dependent self energy by its average over the Brillouin zone; the vanishing of $\mathrm{Im}\Sigma$ for $k>3k_F$ means that the function that is being averaged vanishes over most of integration range so that the DMFT approximation gives an estimate for the low frequency self energy that is  $\mathcal{O}(k_F a)^d$ smaller than  the actual value at the Fermi surface.

A further issue arises for transport: in Galilean and Lorentz-invariant systems the scattering processes that contribute to $\mathrm{Im}\Sigma$ approximately conserve current so the transport scattering rate may be expected to be much smaller than the quasiparticle rate. In diagrammatic calculations the difference between the transport and quasiparticle scattering rates is encoded in vertex corrections which are neglected in practical DMFT calculations. We find that for the low density tight binding model the DMFT underestimation of the quasiparticle rate is to a large degree  compensated by the neglect of the vertex corrections. In Dirac systems additional issues arise that will be discussed below.

The rest of this paper presents the calculations that substantiate the scaling arguments and clarify the issues of vertex corrections and the approach to the low density limit. In Section \ref{sec:low density} we present results for the self energy and low frequency conductivity for tight binding models at low carrier density. In Section \ref{sec:Dirac} we present results for the self energy and low frequency conductivity for the three dimensional Dirac system. Section \ref{sec:conclusion} is a conclusion. Appendices \ref{app: 3D Hubbard} and \ref{app:3D Dirac}
give details of some of the calculations.
 
\section{Low Density Systems}
\label{sec:low density}
\subsection{Formalism}
We analyse the one-band Hubbard model \cite{hubbard1,hubbard2,arovas2022hubbard,qin2022hubbard,leblanc2015solutions,zheng2017stripe,peters2014spin,qin2020absence,xiao2023temperature,wietek2021stripes},
\bea
H=\sum_{\vec{k}\sigma}(E_{\vec{k}}-\mu)c_{\vec{k}\sigma}^{\dagger}c_{\vec{k}\sigma}+U\sum_{i}n_{i\uparrow}n_{i\downarrow},
\eea
with interaction $U$, dispersion $E_k$, lattice constant $a$ and chemical potential $\mu$. We work in units where $\hbar=c=e^2=k_B=1$ throughout the paper and define $\beta$ as the inverse temperature. We are interested in low densities, corresponding to Fermi wave vector $k_F$ small compared to the lattice constant ($k_Fa\ll 1$) or alternatively chemical potential $\mu$ close to the lower band edge ($\mu-min(E_{\vec{k}})\ll W=max(E_{\vec{k}})-min(E_{\vec{k}})$) and work to second order in $U$.

\subsection{Electron self-energy}
The perturbative self energy of interacting electron systems Hubbard models has been extensively studied (see \cite{Hodges71,Fukuyama92,Bloom75,Coffey93,Chubukov06} for some of the original work and references). We recapitulate  the calculations here to obtain expressions in forms convenient for our subsequent analysis. 

The imaginary part of the second order perturbation theory expression for the electron self energy may be expressed in terms of the noninteracting electron spectral function $A(\vec{k},x)=\pi \delta(x+\mu-E_{\vec{k}})$ and the Fermi ($f$) and Bose ($n$) distribution functions as
\begin{eqnarray}
    \mathrm{Im}\Sigma(k,\omega)&=&-U^2a^{2d}\int\frac{d^d\vec{p}_1}{(2\pi)^d}\frac{d^d\vec{p}_2}{(2\pi)^d}\int\frac{dxdy}{\pi^2}A(\vec{p}_1,x)A(\vec{p}_2,y)\nonumber \\ && A(\vec{p}_1+\vec{p}_2-\vec{k},x+y-\omega)
    \left(f(x)-f(x+y-\omega)\right) \nonumber \\ &&\times \left(f(y)+n(y-\omega)\right),
    \label{eq:Imsigma1}
\end{eqnarray}
where $\displaystyle{f(x)=\frac{1}{e^{\beta x}+1}}$ and $\displaystyle{n(x)=\frac{1}{e^{\beta x}-1}}$. The momentum integrals are over the Brillouin zone. At zero temperature, we have $f(x)=\Theta(-x)$ and $n(x)=\Theta(x)-1$, where $\Theta(x)$ is the Heaviside function.  

At temperature $T$ and frequency $\omega$ low with respect to the bandwidth $W$    ($\mathrm{T},\omega \ll W$) and  small momenta ($ka\ll 1$),  the dispersion can be taken to be quadratic, $\displaystyle{E_{\vec{k}}=\frac{k^2}{2m}}$ with $\displaystyle{\mu=\frac{k_F^2}{2m}}$ and as shown in Appendix~\ref{app: 3D Hubbard} Eq.~(\ref{eq:Imsigma1}) can be explicitly written in a scaling form. We present the result here at $T=0$, 
\bea
\mathrm{Im}\Sigma(k,\omega)&=& -\pi(\frac{U}{E_L})^2 (ak_F)^{2d-4} \frac{\mu}{(2\pi)^{2d}}I(\frac{\omega}{\mu},\frac{k}{k_\omega}),
\label{eq:ImSigmaexact}
\eea
where $\displaystyle{E_L=\frac{1}{2ma^2}}$ is a lattice scale energy, the factor $ (ak_F)^{2d-4}$ expresses the scaling of the density of states as the band edge is approached and 
\bea
&&I(\frac{\omega}{\mu},\frac{k}{k_\omega})=\int_1^{1+\frac{\omega}{\mu}}dx x^\frac{d-2}{2} \int_{max[1,1+\frac{\omega}{\mu}-x]}^{2+\frac{\omega}{\mu}-x}dy y^\frac{d-2}{2} \int \frac{d\Omega_x d\Omega_y}{1+\frac{\omega}{\mu}} \nonumber \\ && \times
\delta\Bigg(1+\frac{k^2}{k_\omega^2}+2\frac{\sqrt{x}\sqrt{y}cos\theta_{xy}}{1+\frac{\omega}{\mu}}
-2\frac{\sqrt{x}cos\theta_{xk}+\sqrt{y}cos\theta_{yk}}{\sqrt{1+\frac{\omega}{\mu}}}\frac{k}{k_\omega}\Bigg), \label{eq:I} \nonumber \\
&&k_\omega=\sqrt{k_F^2+2m\omega}=k_F\sqrt{1+\frac{\omega}{\mu}}.
\eea

The rotational invariance of the low energy theory implies $\displaystyle{I(\frac{\omega}{\mu},\frac{k}{k_\omega})}$ only depends on the magnitude of $k$. Eq.~(\ref{eq:ImSigmaexact}) follows directly from scaling the dimensional factors out of the fundamental Eq.~(\ref{eq:Imsigma1}) and expresses the imaginary part of the self energy in the intuitively appealing form of the energy scale of the low energy theory ($\mu$) times $\pi$ times the square of an interaction (made dimensionaless via a lattice scale energy) times the square of a density of states. However for comparison with subsequent work it is convenient to note that in the low density limit of a short ranged interaction model interaction effects are themselves proportional to the particle density (including spin degeneracy) 
$\displaystyle{n=2(ak_F)^d\frac{\Omega_d}{d(2\pi)^d}}$, where $\Omega_d$ is the solid angle of the unit sphere in $d$ dimensions, and to rewrite Eq.~(\ref{eq:ImSigmaexact}) as
\bea
\mathrm{Im}\Sigma(k,\omega)&=& -\pi E_L(\frac{U}{E_L})^2 n\frac{ d}{2(2\pi)^{d}\Omega_d}\left(\frac{\mu}{E_L}\right)^\frac{d-2}{2} I(\frac{\omega}{\mu},\frac{k}{k_\omega}). \nonumber \\
\label{eq:ImSigmaexact1}
\eea


\begin{figure}[h!]
\centering
\subfloat[]{%
  \includegraphics[clip,width=1.0\columnwidth]{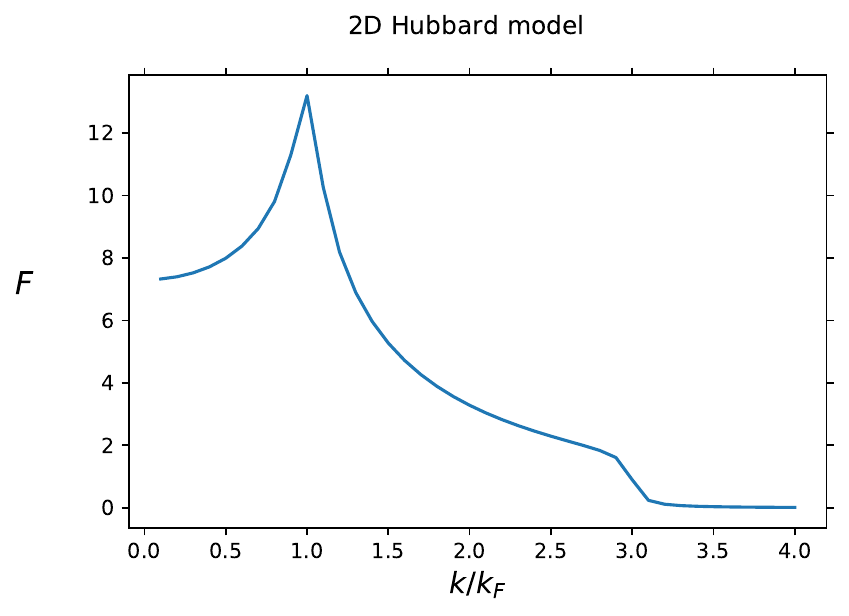}%
}

\subfloat[]{%
  \includegraphics[clip,width=1.0\columnwidth]{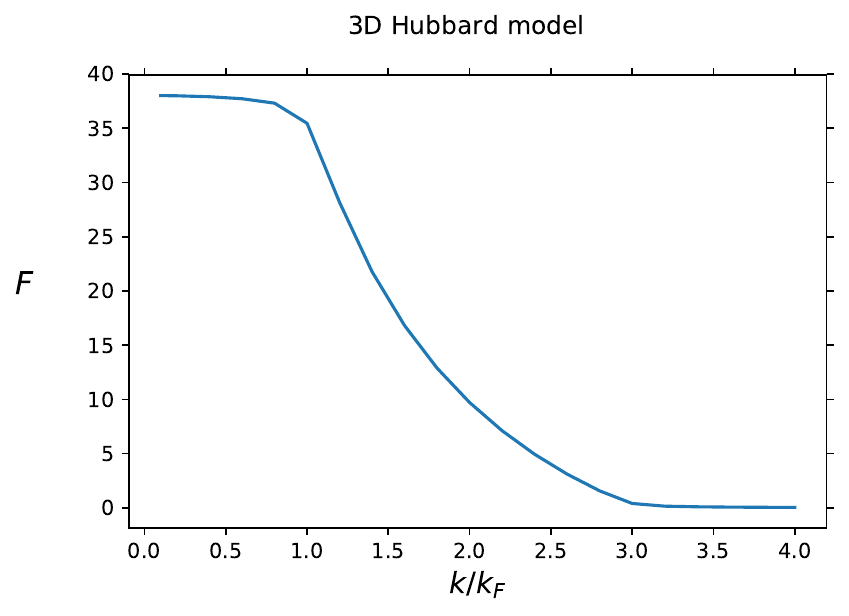}%
}

\caption{ Scaling function $F$ (Eq.~(\ref{eq:F})) shown as a function of $\frac{k}{k_F}$ in two (top panel) and three (bottom panel) dimensions. In the two dimensional case the peak at $k=k_F$ is a logarithmic divergence related to the subleading singularity that gives rise to the $|T|$ correction to the specific heat coeffient extensively discussed in the literature \cite{Coffey93,Chubukov06}. Because the singularity is integrable it does not affect the conductivity. We can clearly see that $F$ vanishes when $k>3k_F$.}
\label{fig:flsigma}
\end{figure}

From Eq.~(\ref{eq:I}) one may easily see that for $\omega/\mu\ll 1$, $x$ and $y$ are almost unity, so the delta function can't be satisfied for $k/k_F>3$, implying that in the low density limit for small $\omega$ the imaginary part of the self energy is zero for most of the Brillouin zone. For large $\omega$ ($W\gg \omega \gg \mu$), $x$ and $y$ can be of order $\omega/\mu$, so the delta function could still be satisfied for $k \sim k_\omega\rightarrow\sqrt{2m\omega}$. Thus as frequency is increased the range of $k$ over which $\mathrm{Im}\Sigma\neq 0$ increases and for large enough $\omega$ the imaginary part of the self energy is nonzero throughout the whole Brillouin zone, although of course the formulas derived above no longer hold.

In the low frequency limit ($\omega \ll \mu$) we have explicitly $k_\omega\rightarrow k_F$ and
\bea
I(\frac{\omega}{\mu}\rightarrow 0,\frac{k}{k_F}) &=& \frac{1}{8}(\frac{\omega}{\mu})^2 F(\frac{k}{k_F}), 
\label{eq:F}
\eea
with $F$ a scaling function given explicitly in Appendix~\ref{app: 3D Hubbard} and plotted for $d=2$ and $3$ in Fig.~\ref{fig:flsigma}. 

\begin{figure}[h!]
\centering
\subfloat[]{%
  \includegraphics[clip,width=1.0\columnwidth]{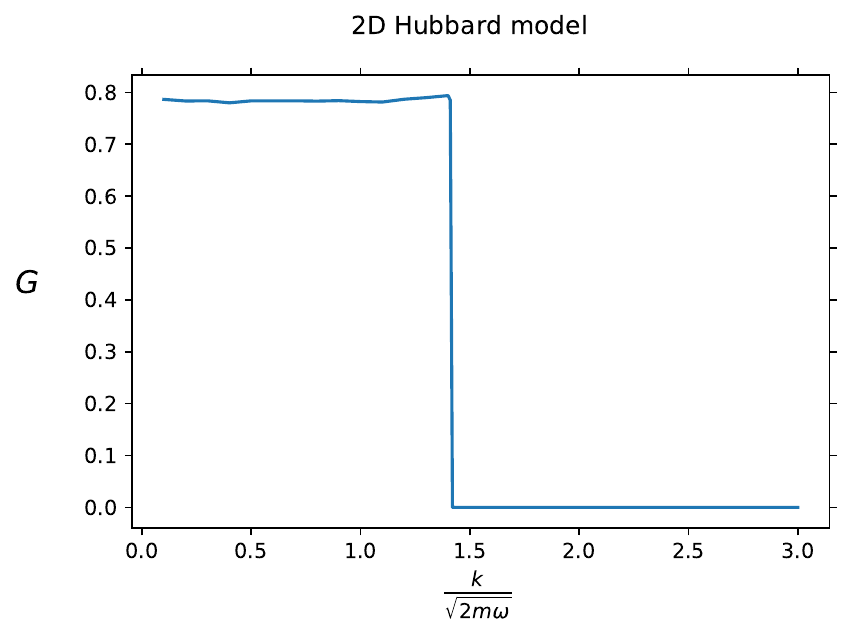}%
}

\subfloat[]{%
  \includegraphics[clip,width=1.0\columnwidth]{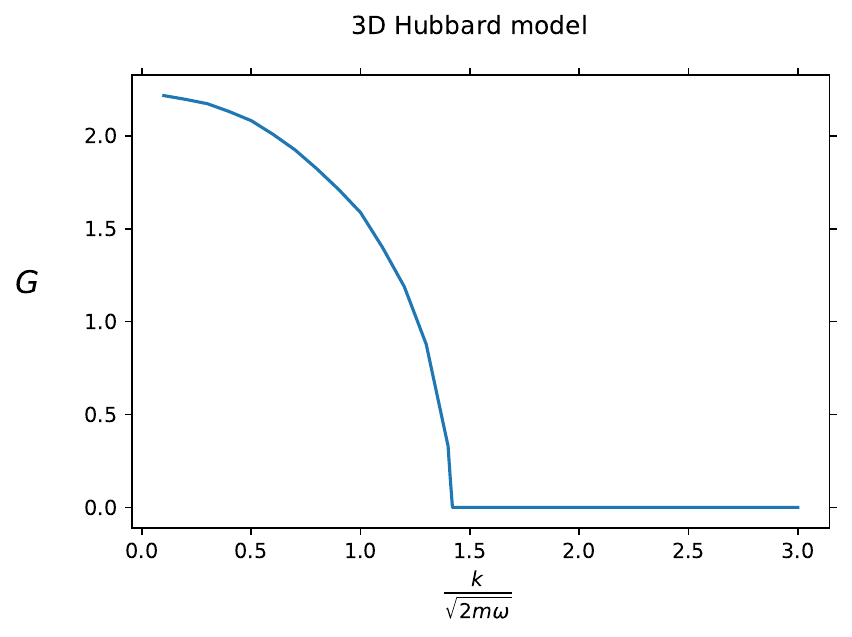}%
}
\caption{ Scaling function $G$ (Eq.~(\ref{eq:Imsigmahlattice})) shown as a function of $\frac{k}{\sqrt{2m\omega}}$ in two (top panel) and three (bottom panel) dimensions. We can clearly see that G vanishes when $\frac{k}{\sqrt{2m\omega}}>\sqrt{2}$. }
\label{fig:hfsigma}
\end{figure}

In the high frequency limit ($W\gg \omega \gg \mu$), $k_\omega\rightarrow \sqrt{2m\omega}$ and we find (see Appendix~\ref{app: 3D Hubbard})
\bea
I(\frac{\omega}{\mu}\gg 1,\frac{k}{k_\omega})&=&
\frac{\Omega_d}{d}(\frac{\omega}{\mu})^{\frac{d-2}{2}}G(\frac{k}{\sqrt{2m\omega}}), 
\label{eq:Imsigmahlattice}
\eea
with $G$ a scaling function given explicitly in Appendix~\ref{app: 3D Hubbard} and plotted for $d=2$ and $3$ in Fig.~\ref{fig:hfsigma}. This scaling means that the self energy becomes independent of $\mu$ and varies with frequency as $\omega^\frac{d-2}{2}$ (of course $\mathrm{Im}\Sigma$ still depends on density via the interaction term $(U/E_L)^2n$). 
From the plot we can see that $G(\frac{k}{\sqrt{2m\omega}})$ goes to a constant for $k<k_\omega=\sqrt{2m\omega}$, so that when $\omega$ is very large we expect $\mathrm{Im}\Sigma(k,\omega)$ to be nonzero over the whole Brillouin zone, although of course the exact expression will differ from Eq.~(\ref{eq:Imsigmahlattice}). 


The second order perturbation theory expression for the dynamical mean field approximation to the self energy is
\begin{eqnarray}
    \mathrm{Im}\Sigma_{\mathrm{DMFT}}(\omega)&=&-U^2a^{2d}\int\frac{d^d\vec{p}_1}{(2\pi)^d}\frac{d^d\vec{p}_2}{(2\pi)^d}\frac{d^d\vec{p}_3}{(2\pi)^d}\int\frac{dxdy}{\pi^2}\nonumber \\ && A(\vec{p}_1,x)
    A(\vec{p}_2,y)A(\vec{p}_3,x+y-\omega)\nonumber \\
&&\times\left(f(x)-f(x+y-\omega)\right) \left(f(y)+n(y-\omega)\right), \nonumber \\
    \label{eq:ImsigmaDMFT1}
\end{eqnarray}
and is seen to be the average, over the Brillouin zone, of the exact perturbative expression implying for $\mu,\omega\ll W$ that
\bea
 \mathrm{Im}\Sigma_{\mathrm{DMFT}}(\omega)&=& -\pi E_L(\frac{U}{E_L})^2 n\frac{ d}{2(2\pi)^{d}\Omega_d}\left(\frac{\mu}{E_L}\right)^\frac{d-2}{2} (ak_\omega)^d \nonumber \\ &\times& \frac{\Omega_d}{(2\pi)^d}\int x^{d-1}dxI(\frac{\omega}{\mu},x). 
\label{eq:ImsigmaDMFT2}
\eea

Since the imaginary part of the perturbative self energy is negligibly small in most of the Brillouin zone (i.e. for $k\gg k_\omega$) we may extend the $x$ integral to infinity. 
 The additional factor of  $(ak_\omega)^d$ shows that at low frequencies ($k_\omega\rightarrow k_F$) the dynamical mean field approximation underestimates the near Fermi surface electron-electron scattering rate by one factor of density; as the frequency is increased the underestimate becomes less severe and for $\omega\sim W$, $k_\omega a\sim 1$ and the DMFT approximation may become reasonable.  
In the low frequency limit, we obtain
\bea
\mathrm{Im}\Sigma_{\mathrm{DMFT}}(\omega)=-(\frac{U}{E_L})^2 (ak_F)^{3d-4} \frac{\pi}{(2\pi)^{3d}}\frac{1}{16}\mu (\frac{\omega}{\mu})^2 (\Omega_d)^3. \nonumber \\
\label{eq:ImsigmaDMFTlowf}
\eea

 We finally consider the quasiparticle weight (mass renormalization), 
\bea
\frac{\partial \mathrm{Re}\sigma(\omega,k_F)}{\partial \omega}\bigg\vert_{\omega\rightarrow 0}&\sim& \mathcal{P}\int\frac{dx}{\pi}\frac{\mathrm{Im}\Sigma(x,k_F)}{(\omega-x)^2} \bigg\vert_{\omega\rightarrow 0}. 
\eea
For $x \ll\mu$, $\mathrm{Im}\Sigma \sim x^2$ while for $x\gg\mu$, $\mathrm{Im}\Sigma\sim x^\frac{d-2}{2}$, so that  for two and three dimensions, the integral is  dominated by $x\sim \mu$; a frequency range where $\mathrm{Im}\Sigma$ is not accurately captured by DMFT. To sum up, in the low density limit of  lattice fermion models, DMFT  parametrically underestimates the low frequency single particle scattering rate and mass enhancement.

\subsection{Optical conductivity}
We now turn to the optical conductivity (response to a long wavelength frequency dependent transverse electric field).  The optical conductivity is determined by the ability of collisions to degrade momentum; in a Galilean-invariant system electron-electron interactions conserve momentum and therefore do not affect the conductivity. In the dilute limit of a lattice model the physics is nearly Galilean-invariant and the effect of interactions on the conductivity is expected to be small. We explore this issue by computing the frequency dependenent conductivity perturbatively to second order in the interaction strength at zero temperature. To understand the calculation it is useful to consider the ``extended Drude" expression for the conductivity,
\begin{equation}
    \sigma(\omega)=\frac{K}{-i(1+\Lambda(\omega))\omega+\Gamma(\omega) }.
    \label{eq:extendeddrude}
\end{equation}
Here $1+\Lambda$ may be thought of as the ``optical mass renormalization" and $\Gamma$ as a frequency dependent optical scattering rate, which in a simple picture of momentum independent scattering is twice the single-particle scattering rate.  $K=\frac{n_{eff}}{m}$ is a basic measure of the density of mobile carriers. In a fermi liquid at low $\omega$ and $T=0$ one expects $\Gamma\sim \gamma_0\omega^2$ while as $\omega\rightarrow 0$ $\Lambda(\omega)$ goes to a constant $\Lambda_0$  so that $lim_{\omega\rightarrow 0}Re[\sigma(\omega)]$ is a constant equal to $\frac{K\gamma_0}{(1+\Lambda_0)^2}$ . In the small interaction limit $\gamma_0,\Lambda_0\propto U^2$  so $\mathrm{Re}\sigma(\omega\rightarrow 0)\rightarrow K\gamma_0$ and in the dilute limit $K=n/m$ where $n$ is the particle density and $m$ is the bare mass (coefficient of the term quadratic in $k$ in the bare dispersion). Comparison of the $\gamma_0$ obtained from the direct calculation of the conductivity to the coefficient of $\omega^2$ in the $\omega\rightarrow 0$ limit of the self energy characterizes the low frequency vertex correction.

We compute the conductivity  perturbatively at zero temperature  via the force-force correlation function method \cite{mahan}. The formalism was described in detail in  \cite{Mu22}. The essential idea is to write the conductivity in terms of the current-current correlation function $\chi_{jj}$ as $\sigma(\omega)=\left(-K+\chi_{jj}(\omega)\right)/i\omega$ and integrate the expression for $\chi_{jj}$  by parts twice, noting that time derivatives correspond to commutators with the Hamiltonian. The current-current correlation function is a tensor in spatial indices $\nu$ and $\nu^\prime$ and on the Matsubara axis
\bea
\chi_{jj}^{\nu,\nu^\prime}(i\omega_n)=-\frac{1}{(i\omega_n)^2}\int_{0}^{\beta}d\tau e^{i\omega_n\tau}\left\langle T_{\tau}F_\nu(\tau)F_{\nu^\prime}(0)\right\rangle, \nonumber \\
\eea
where $F_\nu=\left[H,j_\nu\right]$ is the force operator along direction $\nu$.
 
For the Hubbard model, $\left[H,j_\nu\right]\sim U$ so to leading perturbative order the expectation value in the force-force correlation function may be evaluated at $U=0$. We find for the hypercubic lattice Hubbard model at $T=0$ where $\sigma$ is the unit matrix \cite{Mu22} 
 \bea
\mathrm{Re}\sigma^{\nu,\nu}(\omega)&=&U^2\frac{\pi}{\omega^3}\sum_{\vec{k}\vec{k'}\vec{q}}f(\varepsilon_{\vec{k}})f(\varepsilon_{\vec{k'}})
(1-f(\varepsilon_{\vec{k'}-\vec{q}}))\nonumber \\&&(1-f(\varepsilon_{\vec{k}+\vec{q}})) (v^\nu_{\vec{k}+\vec{q}}+v^\nu_{\vec{k'}-\vec{q}}-v^\nu_{\vec{k}}-v^\nu_{\vec{k'}})^2\nonumber \\
&&\times \delta(\omega+\varepsilon_{\vec{k}}+\varepsilon_{\vec{k'}}-\varepsilon_{\vec{k'}-\vec{q}}-\varepsilon_{\vec{k}+\vec{q}}),
\label{eq:conductivity3d}
\eea
where $\varepsilon_{\vec{k}} = E_{\vec{k}}-\mu$ and  $v^\nu_{\vec{k}}=\partial\varepsilon_{\vec{k}}/\partial k^\nu$. 

We see immediately from Eq.~(\ref{eq:conductivity3d}) that if the energy is quadratic (as is the case in Galilean-invariant or low density  systems) the velocity is linear in $\vec{k}$ and the factor $(v^\nu_{\vec{k}+\vec{q}}+v^\nu_{\vec{k'}-\vec{q}}-v^\nu_{\vec{k}}-v^\nu_{\vec{k'}})^2$ vanishes. The leading low-density contribution is obtained by expanding the velocity beyond quadratic order; because the velocity operator is odd in momentum this means the leading contribution to $(v^\nu_{\vec{k}+\vec{q}}+v^\nu_{\vec{k'}-\vec{q}}-v^\nu_{\vec{k}}-v^\nu_{\vec{k'}})^2$ is $\sim (k_Fa)^6$. We write results specifically for  a cubic lattice tight binding model with dispersion $\varepsilon_{\vec{k}}=-2t(\cos(k_xa)+\cos(k_ya)+\cos(k_za))$.  Expanding the dispersion near $k=0$ identifies the mass as $m=\frac{1}{2ta^2}$. The $z$ direction velocity is $v_{\vec{k}}^z = 2ta\sin(k_za)\approx 2ta(k_za-\frac{(k_za)^3}{6})$. Evaluation of Eq.~(\ref{eq:conductivity3d}) for low frequency $\omega\ll\mu$ yields  $\mathrm{Re}\sigma(\omega)\approx U^2m^2a^6 (k_Fa)^7 \frac{1}{(2\pi)^{9}}{0.24\pi}$ (see Appendix~\ref{app: 3D Hubbard}) per unit cell. It is instructive to express the result as a combination of $n/m$, the single particle scattering rate divided by frequency squared $2/3\mathrm{Im}\Sigma(k_F,\omega)/\omega^2$ (the 2 from the standard relation between the single particle and optical scattering rates and the $1/3$ because in the absence of vertex corrections the conductivity is an average of $\Sigma\sim \omega^2$ over a range of frequencies from $0$ to $\omega$) and the vertex correction

\bea
    \mathrm{Re}\sigma_{perturb}(\omega \ll \mu)&\approx&\frac{n}{m}\frac{2\mathrm{Im}\Sigma(k_F,\omega)}{3\omega^2}\frac{9\times 0.24}{64\pi^3}\left(k_Fa\right)^4  \nonumber \\
    &\sim& \frac{n}{m}(k_Fa)^4.
    \label{eq:sigma3dfinal}
\eea

We see that the vertex correction suppresses the conductivity by a factor $\sim (k_Fa)^4\sim n^\frac{4}{3}$ with a numerical factor which is also very small.

 We now examine the DMFT approximation to the conductivity. Since the self energy is local, only the self energy diagrams remain and vertex corrections vanish (see Appendix \ref{app: 3D Hubbard}). The closed-form expression is obtained in our previous paper \cite{Mu22}
\bea
\mathrm{Re}\sigma^{\nu,\nu}(\omega)&=&U^2\frac{\pi}{\omega^3}\sum_{\vec{k}\vec{k'}\vec{p_1}\vec{p_2}}f(\varepsilon_{\vec{k}})f(\varepsilon_{\vec{k'}})(1-f(\varepsilon_{\vec{p_1}}))\nonumber \\&&(1-f(\varepsilon_{\vec{p_2}})) \Big[{v^{\nu}_{\vec{k}}}^2+{v^{\nu}_{\vec{k'}}}^2+{v^{\nu}_{\vec{p_1}}}^2+{v^{\nu}_{\vec{p_2}}}^2\Big] \nonumber \\
&&\times\delta(\varepsilon_{\vec{k}}+\varepsilon_{\vec{k'}}+\omega-\varepsilon_{\vec{p_1}}-\varepsilon_{\vec{p_2}}).
\label{eq:conductivitydmft}
\eea

Since the integrand in Eq.~(\ref{eq:conductivitydmft}) is always positive, we can just use the quadratic dispersion, and as a result the velocities are of linear order in $ka$. We obtain the low frequency conductivity per unit cell $\mathrm{Re}\sigma(\omega)= U^2m^2a^6 (k_Fa)^6 \frac{(4\pi)^4}{(2\pi)^{12}}\frac{2}{9}\pi.$ This may be written using the low frequency limit of the DMFT scattering rate $\mathrm{Im}\Sigma_{\mathrm{DMFT}}(\omega)=g_{\mathrm{DMFT}}\omega^2$ with $g_{\mathrm{DMFT}}$ extracted from Eq.~(\ref{eq:ImsigmaDMFTlowf}) as 
\begin{equation}
    \mathrm{Re}\sigma_{\mathrm{DMFT}}(\omega\ll\mu)=\frac{n}{m}\frac{2\mathrm{Im}\Sigma_{DMFT}(\omega)}{3\omega^2}\sim \frac{n}{m}(k_Fa)^3.
    \label{eq:sigmadmftfinal}
    \end{equation}  We see that the DMFT underestimate of $\mathrm{Im}\Sigma$ by three factors of $k_Fa$  partially compensates for the neglect of vertex corrections ($k_Fa)^4$. 




\section{Dirac systems}
\label{sec:Dirac}
\subsection{Hamiltonian and formalism}
We now turn to Dirac/Weyl systems and focus on the three dimensional case to provide a straightforward comparison to previous DMFT-based work \cite{Wagner21}.  The new features relative to the previous section arise from the Lorentz, rather than Galilean, invariance of the low energy theory, implying among other things the presence of a low-lying interband transition that breaks current conservation. Also the linear dispersion and low chemical potential means that many features of the electron propagator and self energy are dictated by simple scaling considerations. Finally, at low $T$ and nonzero chemical potential one may consider projecting onto the partially filled band.  As noted in Ref.~\cite{tai2023quantum} this projection is nontrivial.

We follow Ref.~\cite{Wagner21} and consider a lattice model with two  orbitals and two spin states per site and a microscopic (lattice) length scale $a$. At low energies and for momenta near the Dirac point (which we take to be $k=0$), the kinetic (non-interacting) part of the Hamiltonian is
\begin{equation}
H_{kin}=v\sum_{\vec{k}}\Psi_{\vec{k}}^{\dagger}\sigma_z \otimes (\vec{k}\cdot\vec{\tau} )\Psi_{\vec{k}},
\label{eq:Hkin}
\end{equation}
where $\Psi_{\vec{k}}$ is a four-component spinor. $\sigma_z$ and $\vec{\tau}$ are Pauli matrices representing spin and orbital degrees of freedom and the velocity $v$ in combination with the length  $a$ sets the bare energy scales of the model. $H_{kin}$ describes physics at momenta $ka\ll 1$ and energies $\omega$ and temperatures $T$ $\ll v/a$ and we will be interested in lightly doped systems with chemical potentials $0<|\mu| \ll \frac{v}{a}$. The energy dispersion and occupied states are sketched in Fig.~\ref{fig:Dirac}.

In a model with multiple orbitals per unit cell the interaction may take various forms. For ease of notation we take the interaction to be
\begin{equation}
 H_{int}=   U \sum_i\Psi_{i,\uparrow}^{\dagger}\Psi_{i,\uparrow}\Psi_{i,\downarrow}^{\dagger}\Psi_{i,\downarrow},
\end{equation}
where $\Psi_{i,\uparrow}$ is a spin-up two-component spinor describing the orbitals on site $i$ and $\Psi^\dagger\Psi$ denotes  contraction to a scalar in orbital space  so that the interaction is invariant under orbital rotation. Notice that  Ref.~\cite{Wagner21} uses an interaction with no interorbital terms, breaking  the rotation invariance of $H_{kin}$ in orbital space down to a rotation about the z axis.  This difference changes some details of the calculation (in particular the conductivity tensor will not be isotropic, with $\sigma_{zz}\neq \sigma_{xx,yy}$)  but  doesn't affect our qualitative conclusions. 

We define the Green function, a matrix in the spin-orbital basis, in terms of the bare Hamiltonian and the self energy $\Sigma$, also a matrix in the spin-orbital basis, as
\begin{equation}
    G(\vec{k},i\omega_n)=\left(\left(i\omega_n+\mu\right)\mathds{1}-H_{kin}-\Sigma(\vec{k},i\omega_n)\right)^{-1}.
    \label{eq:G}
\end{equation}

Since the interaction is rotationally invariant, the symmetries of the problem dictate that the self energy takes the form 
\begin{equation}
    \Sigma(\vec{k},\omega)=\Sigma_0
    (k,\omega)\sigma_z\otimes \mathds{1}+\Sigma_1 (k,\omega)\sigma_z\otimes \hat{k}\cdot\vec{\tau},
    \label{eq:Sigma}
\end{equation}
where $\hat{k}=\vec{k}/|k|$ is a unit vector parallel to $\vec{k}$. 

The bare Green function $G_0$, given by Eq.~(\ref{eq:G}) with $\Sigma=0$, can be written as a function of general complex frequency argument $z$. We have
\begin{equation}
    G_0(\vec{k},z)=\int\frac{dx}{\pi}\frac{A(\vec{k},x)}{z-x}
    \label{eq:G0spectral},
\end{equation}
where the non-interacting spectral function $A=A_++A_-$ is a matrix in the basis of orbitals with $\pm$ referring to the upper and lower bands respectively and
\begin{equation}
    A_\pm(\vec{k},\omega)=\frac{\pi}{2}\left(\mathds{1}\pm \sigma_z \otimes \hat{k}\cdot\vec{\mathbf{\tau}}\right)\delta(\omega+\mu\mp vk).
    \label{eq:Adef}
\end{equation}

\subsection{Self energy}
To order $U^2$ in the orbital basis the self energy is 
\begin{eqnarray}
    \mathrm{Im}\Sigma^{\uparrow,ab}(\vec{k},\omega)&=&-U^2a^6\int\frac{d^3\vec{p}_1}{(2\pi)^3}\frac{d^3\vec{p_2}}{(2\pi)^3}\int\frac{dxdy}{\pi^2}A^{\uparrow,ab}(\vec{p_1},x)\nonumber \\&& A^{\downarrow,cd}(\vec{p_2},y)A^{\downarrow,dc}(\vec{p_1}+\vec{p_2}-\vec{k},x+y-\omega)
    \nonumber
    \\
    &&\times\left(f(x)-f(x+y-\omega)\right)\left(f(y)+n(y-\omega)\right). \nonumber \\
    \label{eq:Imsigma2}
\end{eqnarray}


We first evaluate Eq.~(\ref{eq:Imsigma2}) at $T=0$ with  chemical potential $\mu>0$ and restrict to frequency $\omega>0$. See Appendix~\ref{app:3D Dirac} for details. The self energy has a scaling form similar to that found for the low density case 
\bea
\mathrm{Im}\Sigma^{\uparrow,ab}(\vec{k},\omega)
    &=&-\pi E_L\frac{U^2}{E_L^2}\frac{(k_\omega a)^5}{(2\pi)^6}\frac{1}{8} I^{ab}\left(\frac{\omega}{\mu},\frac{k}{k_\omega}\right),
    \label{eq:Imsigma3a}
\eea 
with $E_L=v/a$, 
\begin{equation}
    k_\omega=\frac{\omega}{v}+k_F.
    \label{eq:komegadef}
\end{equation}
and $k_F=\mu/v$.

\begin{figure}[h!]
\centering
\subfloat[]{%
  \includegraphics[clip,width=1.0\columnwidth]{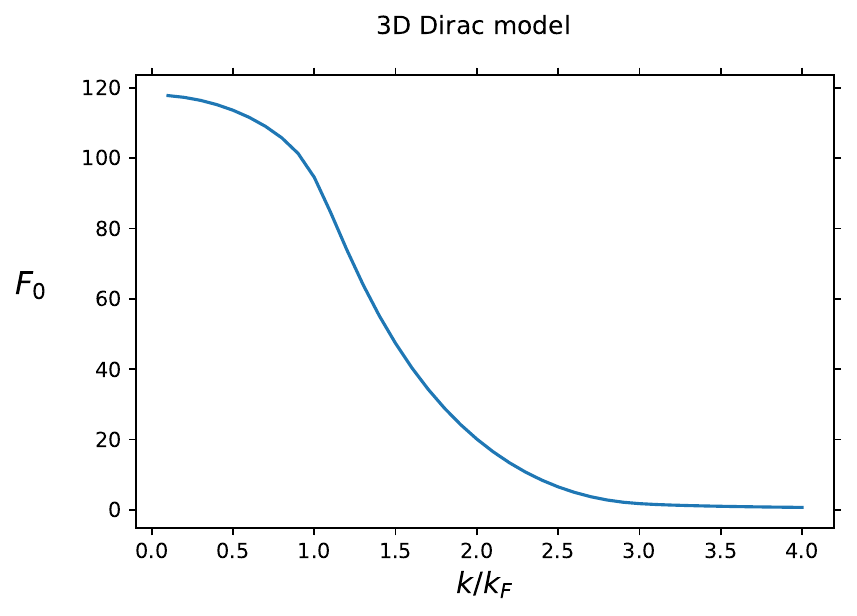}%
}

\subfloat[]{%
  \includegraphics[clip,width=1.0\columnwidth]{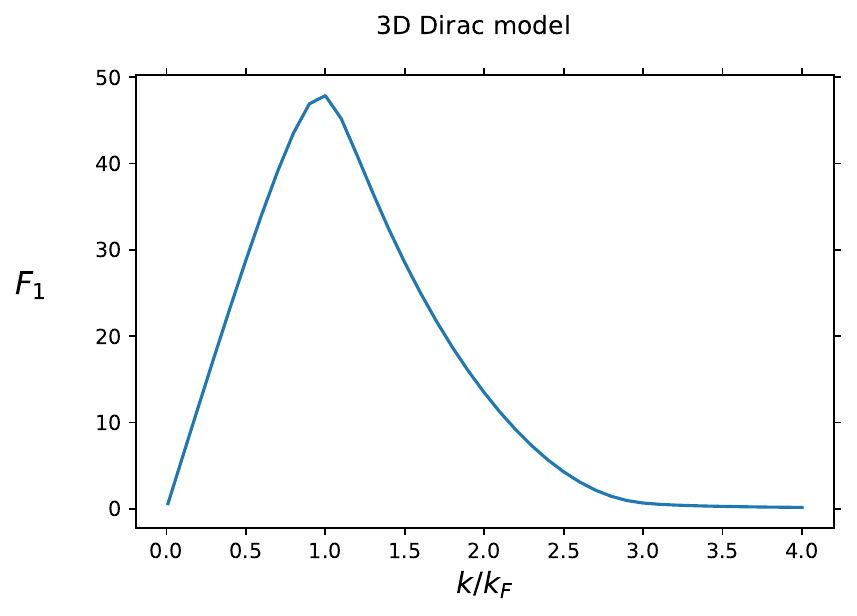}%
}
\caption{The dependence of $F_0$ (panel (a)) and $F_1$ (panel (b)) on $\frac{k}{k_F}$ in three dimension Dirac system. We can clearly see that $F_0$ and $F_1$ vanish when $k>3k_F$.}
\label{fig:3dlfdiracsigma}
\end{figure}

For $\omega\ll\mu$, $k_\omega \rightarrow k_F$ and we have
\begin{equation}
I\left(\frac{\omega}{\mu},\frac{k}{k_F}\right)= (\frac{\omega}{\mu})^2\left(F_0(\frac{k}{k_F})\mathds{1}+F_1(\frac{k}{k_F}) \hat{k}\cdot \vec{\tau}\right), 
\label{eq:Fdiraclow}
\end{equation}
with $F_0$ and $F_1$ are shown in Fig.~\ref{fig:3dlfdiracsigma}. 

\begin{figure}[h!]
\centering
\subfloat[]{%
  \includegraphics[clip,width=1.0\columnwidth]{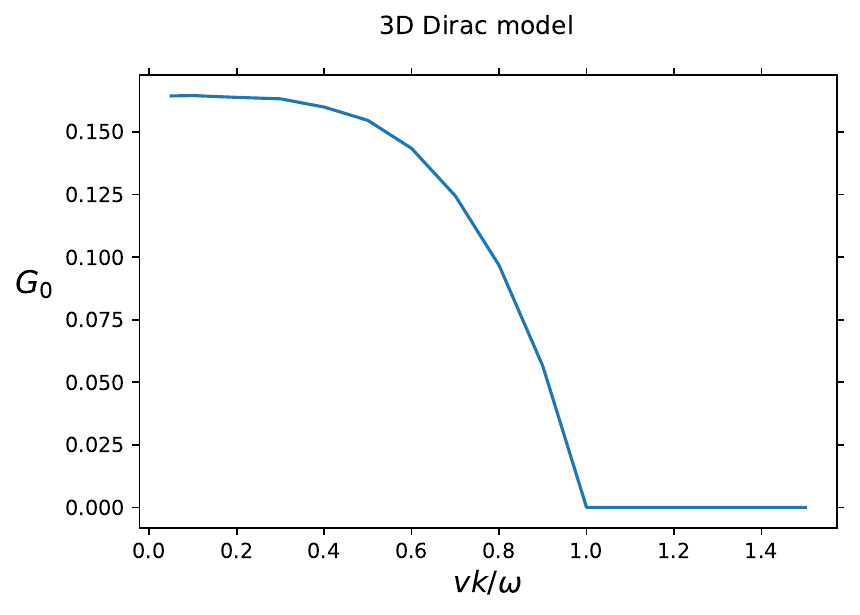}%
}

\subfloat[]{%
  \includegraphics[clip,width=1.0\columnwidth]{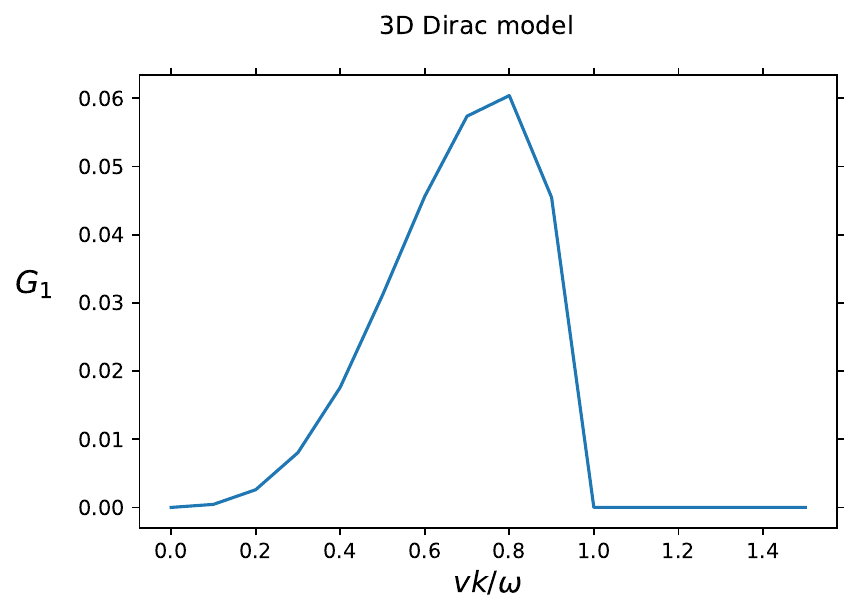}%
}
\caption{The dependence of $G_0$ (panel (a)) and $G_1$ (panel (b)) on $\frac{vk}{\omega}$ in three dimension Dirac system. We can clearly see that $G_0$ and $G_1$ vanish when $vk>\omega$.}
\label{fig:3dhfdiracsigma}
\end{figure}

For $\omega\gg\mu$, $k_\omega \rightarrow \omega/v$ and we have
\begin{equation}
I\left(\frac{\omega}{\mu},\frac{k}{k_\omega}\right) = G_0\left(\frac{k}{\frac{\omega}{v}}\right)\mathds{1}+G_1\left(\frac{k}{\frac{\omega}{v}}\right) \hat{k}\cdot \vec{\tau},
\label{eq:IDirachigh}
\end{equation}
with $G_0$ and $G_1$ shown in Fig.~\ref{fig:3dhfdiracsigma}.


Very similar considerations apply to the case of $T\neq 0$ and $\mu=0$. In terms of dimensionless  variables  we find 
 \begin{equation}
    \mathrm{Im}\Sigma(\vec{k},\omega)=-\pi E_L\frac{U^2}{E_L^2}\frac{(k_{(T,\omega)}a)^5}{(2\pi)^6}\frac{1}{8} I_T\left(\frac{\omega}{T},\frac{k}{k_{(T,\omega)}}\right),
    \label{eq:SigmaT}
    \end{equation}
where  for $\omega \ll T$, $k_{(T,\omega)}\rightarrow k_T=\frac{T}{v}$ and for $\omega \gg T$, $k_{(T,\omega)}\rightarrow k_{\omega}$ (an analytic expression for $k_{(T,\omega)}$ valid for all $T/\omega$ is not available). We do not present the expression for $I_T$ here. As in the previous cases, $\Sigma$ becomes very small for $k\gg k_{(T,\omega)}$, here decaying exponentially.

The content of Eqs.~(\ref{eq:Imsigma3a}),~(\ref{eq:SigmaT}) is that in a low density, low frequency, low temperature  limit the self energy may be written as the product of $\omega^5$, $\mu^5$ or $T^5$ (whichever is largest), and a scaling function of  $\omega/(max(vk,\mu,T))$ and $k/(max(k_F,k_T,k_\omega))$.  The usual Fermi liquid arguments imply that this general structure of the low frequency imaginary part of the self energy applies even beyond the perturbative limit studied here, at least  for frequencies less than $\mu$ and $T$ although the magnitude of the prefactor and the form of the  $k/(max(k_F,k_T,k_{\omega}))$ dependence will depend on the interaction strength. In particular, the vanishing of $\mathrm{Im}\Sigma$ for $k\gg max(k_F,k_T,k_\omega)$ is expected to be general.

We now consider the dynamical mean field approximation. Analogous to the derivation of Eq.~(\ref{eq:ImsigmaDMFT1}), we obtain Eq.~(\ref{eq:ImsigmaDMFTd2}) but with an extra momentum integral:
 \begin{eqnarray}
    &&\mathrm{Im}\Sigma^{\uparrow,ab}_{\mathrm{DMFT}}(\omega)=-U^2a^9\int\frac{d^3\vec{p_1}}{(2\pi)^3}\frac{d^3\vec{p_2}}{(2\pi)^3}\frac{d^3\vec{p_3}}{(2\pi)^3}\int\frac{dxdy} {\pi^2}\nonumber \\ &&A^{\uparrow,ab}(p_1,x)A^{\downarrow,cd}(p_2,y)A^{\downarrow,dc}(p_3,x+y-\omega)
    \nonumber
    \\
    &&\times~\left(f(x)-f(x+y-\omega)\right)\left(f(y)+n(y-\omega)\right).
    \label{eq:ImsigmaDMFTd2}
\end{eqnarray}

The extra momentum integrals lead to the vanishing of the orbital-dependent terms in $A_0$. For $T=0$ and $\mu>0$, when $\omega$ is small,
\begin{equation}
    \mathrm{Im}\Sigma_{\mathrm{DMFT}}^{\uparrow}(\omega,T=0)=-E_L\frac{U^2}{E_L^2}(\frac{\omega}{\mu})^2\frac{(k_Fa)^8}{128\pi^5},
    \label{eq:sigmadmftw}
\end{equation}
while for $T\neq 0$ and $\mu=0$ we obtain
\begin{equation}
    \mathrm{Im}\Sigma^{\uparrow}_{\mathrm{DMFT}}(\omega=0,T)=-\frac{1779.1}{8\pi^5}E_L\frac{U^2}{E_L^2}(k_Ta)^8.
    \label{eq:sigmadmftT}
\end{equation}

The $T^8$ dependence in Eq.~(\ref{eq:sigmadmftT}) is the same power law as obtained by  DMFT calculations in Ref.~\cite{Wagner21}.  

We may understand the difference between the DMFT results Eqs.~(\ref{eq:sigmadmftw}),~(\ref{eq:sigmadmftT}) and the perturbative results, Eq.~(\ref{eq:Imsigma3a}) and Eq.~(\ref{eq:SigmaT}), via the argument that justified Eq.~(\ref{eq:ImsigmaDMFT2}) in the previous section. Since at low frequency the imaginary part of the perturbative self energy vanishes for $k/k_F\gtrsim 1$, by doing the average you get an extra factor $(k_Fa)^3$. Similar arguments hold for the $\mu=0, T\neq 0$ case, where by doing the average you get an extra $(k_Ta)^3$. We can clearly see that DMFT gives wrong exponents compared to the perturbative results. For high frequency, same arguments as in the previous section hold, where DMFT self energy is reasonable.

Turning now to the real part of the self energy, given by
\bea
\mathrm{Re}\Sigma(\omega,k_F)\bigg\vert_{\omega\rightarrow 0}&\sim& \mathcal{P}\int\frac{dx}{\pi}\frac{\mathrm{Im}\Sigma(x,k_F)}{\omega-x} \bigg\vert_{\omega\rightarrow 0} \nonumber \\
&\sim&  \mathcal{P}\int\frac{dx}{\pi}\frac{x^5G(\frac{vk_F}{x})}{\omega-x}.
\eea
Examination of this expression reveals that for the three dimensional Dirac system,  the integral diverges for large $x$ because $G(\frac{vk_F}{x})$ approaches a constant for $\frac{vk_F}{x}\rightarrow 0$. The divergence is cut off when $x$ reaches a lattice scale energy $v/a$. This means that  the real part of the self energy and therefore the mass enhancement are dominated by $\mathrm{Im}\Sigma$ at lattice scales, where we expect the self energy to have only relatively weak momentum dependence and therefore to be reasonably approximated by DMFT.

To sum up, in the $3D$ Dirac system the DMFT approximation for the imaginary part of the low frequency self energy is incorrect because it strongly underestimates low frequency near Fermi surface single particle scattering rate, but may be a reasonable approximation to the real part of the self energy and therefore may  get issues like the location of the Mott transition right.

\subsection{Optical conductivity}
We evaluate the conductivity  at zero temperature and nonzero $\mu$, to second order in the interaction strength, using the force-force correlation function formalism. The optical conductivity of Dirac systems has additional features relative to the tight binding systems because  the Lorentz, rather than Galilean, invariance of the low energy theory  among other things means that  a low energy (threshold $2\mu$) interband transition leads to current non-conservation although momentum is conserved.  We focus on the low frequency limit $\omega\ll\mu$ so that real interband transitions are not relevant.  

We consider the z direction current, which is given by $\displaystyle{j_z=\frac{\partial H_{kin}}{\partial k_z}=v\sum_{\vec{k}}\Psi_{\vec{k}}^{\dagger}\sigma_z\otimes \tau_z}\Psi_{\vec{k}}$. Our  rotationally invariant $H_{int}$ commutes with $j_z$: current non-conservation arises wholly from the interband transitions in the non-interacting part so the force operator is
\bea
F_z=[H_{kin},j_z] = 2iv^2\sum_{\vec{k},\gamma}\Psi_{\vec{k},\gamma}^{\dagger}(\vec{\tau} \times \vec{k})\cdot \hat{z}\Psi_{\vec{k},\gamma},
\eea
where $\gamma$ is the spin index. In the energy basis that diagonalizes $H_{kin}$, $F_z$ has only off-diagoanl matrix elements, reflecting the role of interband transitions in breaking current conservation. Unlike in the low density tight binding model case, where the force operator was proportional to the interaction and the force-force correlation function could be evaluated in the non-interacting limit, here the force-force correlation function must be evaulated to seconrd order in the interaction strength. 

\begin{figure}[h!]
\centering
\subfloat[type one]{%
  \includegraphics[clip,width=0.7\columnwidth]{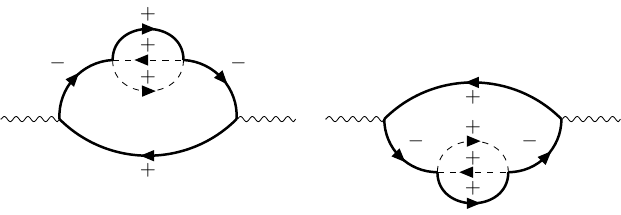}%
}

\subfloat[type two]{%
  \includegraphics[clip,width=0.7\columnwidth]{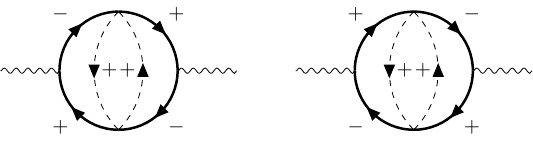}%
}

\subfloat[type three]{%
  \includegraphics[clip,width=0.7\columnwidth]{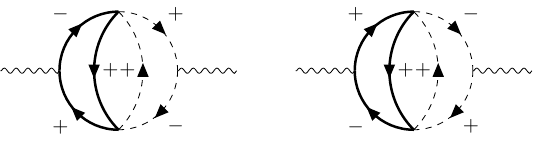}%
}

\subfloat[type four]{%
  \includegraphics[clip,width=0.7\columnwidth]{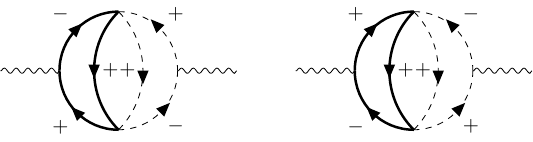}%
}

\caption{All second order diagrams contributing to the low frequency optical absorption. Here type one is the self energy diagrams and type two, three and four are the vertex diagrams. $+$ correpsonds to top band while $-$ corresponds to lower band. In panel (a), (b), (c) and (d), the dashed and thick lines indicate the paths over which orbital indices are traced. Each diagram has a spin degeneracy of two.}
\label{fig:diagrams}

\end{figure}

 Constructing the force-force correlation function as before to second order in $U$ we find the eight diagrams shown in Fig.~\ref{fig:diagrams} and representing  an energy dissipation process that involves photon absorption via a virtual interband transition followed by an interaction-induced downscattering to fill the hole in the lower band. In each diagram there are six fermion lines; the band off-diagonal nature of $F$ means that at least two of the fermion lines lie in the lower (filled)  bands. At $|\omega|<\mu$  obtaining a non-zero conductivity requires that all four remaining lines lie in the upper bands and the diagrams are distinguished in part by the assignment of band indices to the Green function lines. In drawing the diagrams it is also convenient to distinguish  different spins (solid and dashed lines) because at the interaction vertex the incoming and outgoing line of the same spin also have the same orbital index.     The diagrams  may be organized into self energy diagrams (``type one", panel a), and three kinds of vertex correction diagrams (types 2-4, panels b-d) distinguished by the nature of the inter/intra-orbital scattering  and by whether the vertex correction involves a particle-hole (types two and three)  or particle-particle (type 4) pair.   
 
\begin{figure}[h!]
\centering
\subfloat[\label{fig:Dirac}]{%
  \includegraphics[clip,width=0.5\columnwidth]{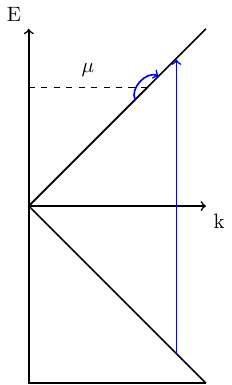}%
}

\subfloat[\label{fig:roughplot}]{%
  \includegraphics[clip,width=1.0\columnwidth]{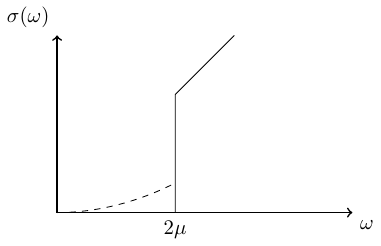}%
}   
\caption{(a) Sketch of dispersion showing filled states below chemical potential (dashed line). The arrows represent intraband and interband transitions. (b) Sketch of optical conductivity of 3D Dirac system. Dashed line represents intraband transition and solid line represents interband transition.}
\label{fig:two graphs}
\end{figure}

 The diagrams are evaluated in  Appendix \ref{app:3D Dirac}. Each diagram gives a contribution to the optical conductivity that is a constant  plus terms of higher power in frequency. The constant contribution of a diagram of type j is $c_j\frac{1}{(2\pi)^9}\frac{1}{2^6}\frac{4096}{135}\pi^4\frac{U^2}{(\frac{v}{a})^2}(k_Fa)^5a^2$  per unit cell. We find   $c_1=1$, $c_2=-\frac{3}{7}$ and $c_3=c_4=-\frac{2}{7}$ so that  $c_1+c_2+c_3+c_4=0$ and the diagrams exactly cancel  implying that the low frequency optical conductivity approaches zero as $\omega\rightarrow 0$, with the first correction $\sim \omega^2$, as shown in Fig.~\ref{fig:roughplot}, implying a transport scattering rate $\sim \omega^4$.

An alternate approach to computing the  conductivity at $|\omega|\ll\mu$ is to project the Hamiltonian onto the partially occupied band. The current operator becomes the velocity in the relevant direction and the interaction acquires a non-trivial momentum dependence. The topological/geometric phase considerations noted in \cite{tai2023quantum} are higher order in the interaction amd frequency.  Applying the force-force method to this effective one band model also leads to a conductivity $\sim \omega^2$. 


The dynamical mean field approximation to the conductivity is subtle. In the one band tight binding model the vectorial nature of the current operator means that in the DMFT approximation the current vertex vanishes, so the conductivity may be evaluated just from the self energy diagrams. However, in the Dirac system, the current operator  $j_z\sim \tau_z$ has interband terms that are not odd in momentum, which would be subject to vertex corrections even in a momentum-independent self energy approximation. As noted above, the projection onto the conduction band leads to an interaction with an explicit momentum dependence that is important in the calculation of the current-current correlation function. Thus even within a  momentum-independent self energy approximation, properly defining a DMFT approximation requires care. However in general in this approximation one would expect that $Re[\sigma(\omega\rightarrow 0)]\neq 0$.  For example, evaluating the current-current correlation function within the DMFT spirit by retaining on the self energy diagram (type 1) and evaluating the self energy in the DMFT approximation via Eq.~(\ref{eq:sigmadmftw}) yields $\displaystyle{\mathrm{Re}\sigma(\omega)=\frac{1}{(2\pi)^3}\frac{1}{128\pi^5}\frac{16\pi}{9}\frac{U^2}{(\frac{v}{a})^2}(k_Fa)^8 }a^2$ per unit cell, which is non-vanishing, in contrast to the exact answer. Notice that this expression can again be written as $\mathrm{Re}\sigma(\omega\ll \mu)=K\frac{2\mathrm{Im}\Sigma_{DMFT}(\omega)}{3\omega^2}$, where $K=2\frac{\mu^2}{6\pi^2v}$. $\frac{\mu^2}{6\pi^2v}$ is the weight of a single node in the $3D$ Dirac case, $2$ corresponds to the spin degeneracy and $\mathrm{Im}\Sigma_{\mathrm{DMFT}}(\omega)$ follows Eq.~(\ref{eq:sigmadmftw}).

\section{Summary}
\label{sec:conclusion}

It is  well understood that the low density and low energy limit of a lattice model will in general be described by a translation invariant theory with subleading corrections. However the implications of this fact for the dynamical mean field theory description of correlated electron physics seem not to have been fully discussed. In this paper we have addressed this issue by combining a perturbative approach and scaling analysis to compute both the exact (perturbative) and DMFT approximations to the  self energy and $T=0$ low frequency conductivity of two paradigmatic systems: the Hubbard model and a Dirac/Weyl metal system. In both cases there is a fundamental Hamiltonian which is described by a microscopic length (lattice constant) $a$ and at temperature $T=0$ the electron density may be expressed in terms of a Fermi wave vector $k_F$ and we are interested in the limit $k_Fa\rightarrow 0$ which corresponds to a nearly empty band and emergent Galilean invariant  in the Hubbard model case and a chemical potential near the Dirac point and emergent Lorentz invariance in the Dirac case.   

We find that in both the Hubbard and Dirac cases  the DMFT approximation provides a qualitatively inaccurate representation of the low frequency scattering rate, too small by a factor proportional to the particle density $(k_Fa)^d$. The essential reason, which indicates that our arguments hold generally beyond the perturbative limit studied in detail here, is that in the low density/low frequency limit the only relevant scale is the fermi wavevector $k_F$, so the imaginary part of the self energy is a function of $k/k_F$ that vanishes when $k/k_F$ becomes larger than a number of the order of unity. This implies a strong momentum dependence of the self energy that is incompatible with the DMFT approximation, which in a d-dimensional system underestimates the near fermi surface scattering rate by a factor $\propto (k_Fa)^d$, in other words proportional to the density. As the frequency $\omega$ becomes larger than the chemical potential, the imaginary part of the self energy becomes a scaling function of $k/k_\omega$ where the momentum scale $k_\omega$ increases with $\omega$ and eventually becomes comparable to the lattice scale $a^{-1}$, by which point the DMFT approximation may be expected to be qualitatively reasonable. We find that for the Dirac, but not the low density Hubbard systems, the real part of the self energy is dominated by high frequency processes such that $k_\omega a\sim 1$, so the real part of the self energy of the Dirac system may be correctly described within DMFT.

We also used the force-force correlation function method to study the optical conductivity at $T=0$ for frequencies less than the chemical potential. This method provides a compact and convenient representation of the vertex corrections which represent the difference between scattering process which degrade the current and those that do not. In a fermion system at $T=0$ and frequency $\omega\ll\mu$ the single particle scattering rate can be represented as $g_{sp}\omega^2/\mu$ while the conductivity can be represented in terms of an optical scattering rate $g_{opt}\omega^2/\mu$. In the dynamical mean field approximation $g_{opt}=2g_{sp}$ while in a strictly Galilean-invariant system the identity of the current and momentum operators means  $g_{opt}=0$. In the low density d-dimensional Hubbard system we find $g_{sp}\sim \left((k_Fa)^{d-2}\right)^2$ with the factors $(k_Fa)^{d-2}$ representing the density of states which vanishes at the band edge for $d>2$. In 3d $g_{opt}\sim g_{sp}(k_Fa)^4$, with the four extra factors reflecting the subleading corrections that break the Galilean invariance of the low energy theory. In $d=2$ the interplay of scattering kinematics and momentum conservation means that $g_{opt}=0$ for $k_Fa$ less than a critical value corresponding to the threshold for Umklapp scattering \cite{Mu22}. In the Dirac case the model interaction used here commutes with the current but real interband transitions occurring at $\omega>2\mu$ break current conservation. Some care is thus required in the computation of the conductivity but we find  that $g_{opt}=0$ with corrections of order $\omega^2$.  The dynamical mean field approximation, which does not include vertex corrections,  gives an incorrect representation of the conductivity. 

Our results raise questions regarding recent DMFT-based calculations of transport in Weyl systems \cite{Wagner21,Poli23}. These papers report a $T^6$  resistivity for a 3D Dirac system at the particle-hole symmetric point from the combination of the DMFT-predicted $T^8$ scattering rate and a $T^2$  Drude weight (or effective carrier density).  Our results indicate that the DMFT results for the imaginary part of the self energy and for the transport scattering rates \cite{Wagner21,Poli23} are not correct for the low frequencies and temperatures where Dirac points dominate the physics, and suggest (although we did not consider the $\omega=0$, $T\neq0$ conductivity explicitly), that vertex corrections also need to be considered.  The issue of interaction corrections to the conductivity of Weyl systems should be revisited, building on previous quantum Boltzmann equation work \cite{Lars08,Maslov11,Pal12} . In this context it is also interesting to consider an orbitally anisotropic Hamiltonian, since many physically relevant  Weyl metals have multiple Dirac points related by symmetry so that the Hamiltonian for one such point may not have the full symmetry of the crystal.   

Further, a number of interesting $d^1$ transition metal oxides such as $SrVO_3$ or $SrMoO_3$ may be viewed as being in a low density limit since the density of electrons per active spin-orbital state is $1/6$. While the Fermi surfaces are complicated, comparison of perturbative and DMFT calculations should be performed to confirm the accuracy of DMFT in these systems.  

On the theoretical side, in the version used here, the force-force correlation is a high frequency expansion. At $T=0$ in a clean Fermi liquid at weak coupling the results can be applied down to $\omega=0$ but alternative treatments of vertex corrections that would remove the restrictions of weak coupling and $T=0$ would be of great interest.  

\begin{acknowledgments}
The authors acknowledge helpful discussions with G. Sangiovanni and N. Wagner and  support from Programmable Quantum Materials, an Energy Frontier Research Center funded by the U.S. Department of Energy (DOE), Office of Science, Basic Energy Sciences (BES), under award DE-SC0019443.   The Flatiron Institute is a division of the Simons foundation.
\end{acknowledgments}

\appendix
\begin{widetext}
\section{low density 3D Hubbard model}
\label{app: 3D Hubbard}
\subsection{Self Energy}
This subsection gives some details of the perturbative self energy calculation for the low density limit of the Hubbard model. Introducing a lattice-scale energy
\begin{equation}
E_L=\frac{1}{2ma^2},
\label{eq:EL}
\end{equation}
and a single-spin density of states 
\begin{equation}
N(\varepsilon)=a^d\int p^{d-1}dp~\delta\left(\frac{p^2}{2m}-\varepsilon\right)=\frac{1}{2E_L}\left(\frac{\varepsilon}{E_L}\right)^\frac{d-2}{2},
\label{eq:N}
\end{equation}
and taking into account the limits of integration imposed by the distribution functions and scaling $x$ and $y$ by $\mu$ and then shifting them by $1$ we may express Eq.~(\ref{eq:Imsigma1})  at $T=0$ as 
\begin{equation}
\mathrm{Im}\Sigma(\omega,k)=-\pi E_L\frac{U^2}{E_L^2}\frac{nd}{8(2\pi)^{d}\Omega_d}\left(\frac{\mu}{E_L}\right)^\frac{d-2}{2}\int_1^{1+\frac{\omega}{\mu}}dx x^\frac{d-2}{2} \int_{max[1,1+\frac{\omega}{\mu}-x]}^{2+\frac{\omega}{\mu}-x}dy y^\frac{d-2}{2}\mathcal{I}\left(\frac{\omega}{\mu},\frac{k}{k_\omega},x,y\right). 
\label{eq:sigmascaled}
\end{equation}
with
\begin{eqnarray}
    \mathcal{I}&=&\int \frac{d\Omega_x d\Omega_y}{1+\frac{\omega}{\mu}}\delta\left(1+\frac{k^2}{k_\omega^2}+2\frac{\sqrt{x}\sqrt{y}cos\theta_{xy}}{1+\frac{\omega}{\mu}}-2\frac{\sqrt{x}cos\theta_{xk}+\sqrt{y}cos\theta_{yk}}{\sqrt{1+\frac{\omega}{\mu}}}\frac{k}{k_\omega}\right),
    \label{eq:Idef} 
\end{eqnarray}
where $n=2(k_Fa)^d \frac{\Omega_d}{d(2\pi)^d}$ is the particle density (including spin degeneracy), $\Omega_{x,y}$ are the solid angles describing the orientation of $\vec{p}_1$ (magnitude $k_F\sqrt{x}$) and $\vec{p}_2$ (magnitude $k_F\sqrt{y}$) on the unit sphere in $d$ dimensions, $\theta_{xy}$ is the angle between $p_1$ and $p_2$ etc. and 
\begin{equation}    k_\omega=\sqrt{k_F^2+2m\omega}=k_F\sqrt{1+\frac{\omega}{\mu}}.
\end{equation}
Here the limits on the $x$ and $y$ integrals express the requirements that for $\omega>0$ $\varepsilon_{p_1},\varepsilon_{p_2}>0$ while $-\mu<\varepsilon_{p_1+p_2-k}<0$.

Eq.~(\ref{eq:sigmascaled}) makes it manifest that in the low frequency and density limits $\mathrm{Im}\Sigma$ has a magnitude determined by the square of the interaction (made dimensionless by a lattice scale energy) times the particle density (because for short ranged interactions all interaction effects must scale with the density) and its frequency and momentum dependence is given as a scaling function of the frequency normalized to the chemical potential and momentum normalized to $k_\omega$ and (for dimensions above the marginal dimension $d=2$) a factor of a low frequency scale normalized to the lattice scale. While the precise form of the scaling function and definition of the lattice scale will change if the calculation is pushed to higher orders in the interaction, we expect that this qualitative behavior is generic.

In the low frequency limit, we may set  $x=1+u$ and $y=1+v$ and neglect $\omega,u,v$ in the expression for $\mathcal{I}$ and perform the $u$ and $v$ integrals to leading order in $\omega/\mu$. Defining $F=\mathcal{I}\left(0,\frac{k}{k_F},1,1\right)$  we obtain
\begin{equation}
    \mathrm{Im}\Sigma(\omega,k)=-\pi E_L\frac{U^2}{E_L^2}\frac{nd}{8(2\pi)^{d}\Omega_d}\left(\frac{\mu}{E_L}\right)^\frac{d-2}{2}\frac{\omega^2}{2\mu^2}F\left(\frac{k}{k_F}\right).
    \label{eq:sigmalowfinal}
\end{equation}


In the two dimensional case, due to rotational invariance we can choose the x direction of the internal coordinates to be parallel to $\vec{k}$ obtaining
\bea
F(x) = \int d\theta_1 d\theta_2 \delta(-1-2\cos(\theta_1)\cos(\theta_2)-2\sin(\theta_1)\sin(\theta_2)+2x(\cos(\theta_1)+\cos(\theta_2))-x^2), 
\eea
where $\theta_{1,2}$ are the angles between $x$ axis and $\vec{p}_{1,2}$.

For three dimensions, we can choose the z axis of the internal coordinates to be parallel to $\vec{k}$  obtaining
\bea
F(x) &=& 2\pi\int d\theta_1 d\theta_2 d\phi~\sin(\theta_1)\sin(\theta_2)\\
&\times&\delta\left((-1-2\cos(\theta_1)\cos(\theta_2)-2\sin(\theta_1)\sin(\theta_2)\cos(\phi) 
+2x(\cos(\theta_1)+\cos(\theta_2))
-x^2\right), 
\nonumber
\eea
where $\phi$ is the azimuthal angle difference between $\vec{p_1}$ and $\vec{p_2}$, and $\theta_{1,2}$ are the polar angle between $z$ axis and $\vec{p}_{1,2}$.

We now look at the high frequency limit $\omega\gg\mu$ but $\omega\ll W$.  While it is possible to proceed from the scaling form Eq.~(\ref{eq:Idef}) it is more convenient to begin from the fundamental equation, which may be written as
\bea
\mathrm{Im}\Sigma(k,\omega)&=&-U^2a^{2d}\int\frac{d^d\vec{p_1}}{(2\pi)^d}\frac{d^d\vec{p_2}}{(2\pi)^d} \pi \delta(E_{\vec{p}_1+\vec{p_2}-\vec{k}}+\omega-E_{\vec{p}_1}-E_{\vec{p}_2}+\mu) \nonumber \\
&\times&(f(E_{\vec{p}_1+\vec{p_2}-\vec{k}}-\mu)(1-f(E_{\vec{p}_1}-\mu)(1-f(E_{\vec{p}_2}-\mu)) \nonumber \\
&+&(1-f(E_{\vec{p}_1+\vec{p_2}-\vec{k}}-\mu))f(E_{\vec{p}_1}-\mu)f(E_{\vec{p}_2}-\mu)),
\label{eq:rewrite}
\eea
with $E_p=\frac{p^2}{2m}$.

For $\omega>0$, at $T=0$, only the first term in the sum survives and we may change variables to $q=p_1+p_2-k$ and $p_1$. In the high $\omega$ limit we may set $E_q=0$ and $\mu=0$ inside the $\delta$ and omit the $q$  in the definition of $p_2$, perform the integral over $q$ trivially, obtaining
\bea
\mathrm{Im}\Sigma(k,\omega)&=&-\pi U^2\frac{a^{d}n}{2(2\pi)^{d}}\int d^d\vec{p_1} \delta\left(\omega-\frac{p_1^2}{2m}-\frac{\left(\vec{k}-\vec{p}_1\right)^2}{2m}\right).  
\label{eq:rewrite1}
\eea
Now scaling the $\omega$ out of the $\delta$, recalling the high frequency limit $k_\omega\rightarrow \sqrt{2m\omega}$ we obtain
\bea
\mathrm{Im}\Sigma(k,\omega)&=&-\pi E_L \frac{U^2}{E_L^2}\frac{n}{2(2\pi)^{d}}\left(\frac{\omega}{E_L}\right)^\frac{d-2}{2}G\left(\frac{k}{k_\omega}\right), 
\label{eq:rewrite2}
\eea
with (writing $G$ explicitly in two and three dimensions)
\begin{eqnarray}
G_2(x)&=&\int dv v \int d\theta \delta\left(1-2v^2-x^2+2xv cos(\theta)\right),
\label{eq:G2def}
\\
G_3(x)&=&2\pi\int dv v^2 \int sin\theta d\theta \delta\left(1-2v^2-x^2+2xvcos(\theta)\right).
\label{eq:G3def}
\end{eqnarray}


\subsection{Conductivity}

Here we present explicit formulas for the conductivity for the three dimensional case (the two dimensional case, which presents some special features,  was discussed in Ref ~\cite{Mu22}), and for $\omega\ll\mu$. We use the perturbative force-force correlation approach and without loss of generality focus on the z-z conductivity.

It is  convenient to rewrite the diagonal components of the conductivity (Eq.~(\ref{eq:conductivity3d})) as 
\bea
\mathrm{Re}\sigma^{aa}(\Omega)&=&\frac{2U^2}{\pi\Omega^3}\sum_{\vec{q}}\int_{-\Omega}^0 d\omega \bigg[B_a^{(2)}(\vec{q},\omega+\Omega)B_a^{(0)}(-\vec{q},-\omega) \nonumber \\
&+&B^{(1)}(\vec{q},\omega+\Omega)B_a^{(1)}(-\vec{q},-\omega)\bigg],
\label{eq:twobubbles}
\eea
where 
\bea
B_a^{\alpha}(\vec{q},\omega)&=&-\pi\sum_{\vec{k}}(f(\varepsilon_{\vec{k}})-f(\varepsilon_{\vec{k}}+\omega))\delta(\omega+\varepsilon_{\vec{k}}-\varepsilon_{\vec{k}+\vec{q}})(v^a_{\vec{k}+\vec{q}}-v^a_{\vec{k}})^\alpha, \nonumber \\
\alpha&=&0,1,2.
\label{eq:Balpha}
\eea
We assume cubic symmetry so $\mathbf{\sigma}\sim \mathbf{1}$ and without loss of generality we may choose $a=z$. Inspection of Eq.~(\ref{eq:conductivity3d}) shows that we need to retain only the terms in $(v^a_{\vec{k}+\vec{q}}-v^a_{\vec{k}})$ cubic in wave vector, so $v^z_{\vec{k}+\vec{q}}-v^z_{\vec{k}}=-\frac{ta^4}{3}\left(3k_z^2q_z+3k_zq_z^2+q_z^3\right)$ where we used $t=1/2ma^2=E_L$.

We see that at low frequency and $T=0$ the combination of Fermi function and delta function constrains $k$ and $k+q$ to lie very near to the Fermi surface, implying a relation between the magnitude of $q$ and the angle between $k$ and $q$ that is most conveniently treated using coordinates for $k$ that are tied to the direction of $q$. We adopt these coordinates and  therefore write $k_z=kcos\theta_{k\hat{z}}$ and view  $cos\theta_{k\hat{z}}$ as a function of the angular coordinate  specifying the orientation of $k$ with respect to  $q$ and the orientation of $q$ with respect to $\hat{z}$.  At small $\omega$ we may set $|\vec{k}|=k_F$ and scale $q$ by $k_F$ obtaining
\begin{equation}
B^\alpha(\vec{q},\omega)=-\frac{\pi}{(2\pi)^3}(-\frac{1}{3})^{\alpha}m^2\omega t^{\alpha}(k_Fa)^{3\alpha-1}a^{\alpha+4}\frac{k_F}{q}I_\alpha(\frac{q}{k_F},cos\theta_q),
\label{eq:Bscaled}
\end{equation}
with
\begin{equation}
I_\alpha(\bar{q},cos\theta_q)=\int d\phi_{kq}\left(3\bar{q}cos^2\theta_{k\hat{z}}cos\theta_q+3\bar{q}^2cos\theta_{k\hat{z}}cos^2\theta_q+\bar{q}^3cos^3\theta_q\right)^\alpha,
\label{eq:Ialphadef}
\end{equation}
where implicit in the definition of $I$ and $B$ is that $cos\theta_{kq}=-\frac{\bar{q}}{2k_F}$ and that $0<\bar{q}<2$. $\theta_q$ is the angle between $\vec{q}$ and $\hat{z}$, and $\theta_{kq}$ is the angle between $\vec{q}$ and $\vec{k}$.

To obtain the expressions for $\cos\theta_{k\hat{z}}$, we introduce the rotation matrix
\bea
R(\theta,\phi)=\begin{bmatrix}
\cos(\theta)+\sin^2(\phi)(1-\cos(\theta)) & -\sin(\phi)\cos(\phi)(1-\cos(\theta)) & \sin(\theta)\cos(\phi)\\
-\sin(\phi)\cos(\phi)(1-\cos(\theta)) &\cos(\theta)+\cos^2(\phi)(1-\cos(\theta))& \sin(\theta)\sin(\phi)\\
-\sin(\theta)\cos(\phi)&-\sin(\theta)\sin(\phi)&\cos(\theta) 
\end{bmatrix}. 
\eea

When $R$ acts on a vector $(0,0,q)$, it will give the vector $q(\sin(\theta)\cos(\phi),\sin(\theta)\sin(\phi),\cos(\theta))$.

In the frame of $q$ the coordinates of $k$ is $k(\sin(\theta_{kq})\cos(\phi_{kq}),\sin(\theta_{kq})\sin(\phi_{kq}),\cos(\theta_{kq})).$ To calculate $k_z$, we use $R(\theta_q,\phi_q)$ to act act on this vector, and then project on the z axis. We then have
\bea
\cos\theta_{k\hat{z}} = -\sin(\theta_q)\cos(\phi_q)\sin(\theta_{kq})\cos(\phi_{kq})-\sin(\theta_q)\sin(\phi_q)\sin(\theta_{kq})\sin(\phi_{kq})+\cos(\theta_q)\cos(\theta_{kq}).\nonumber \\
\eea

Combining the expressions and performing the convolutions  gives
\begin{equation}
    \sigma=\frac{U^2\pi m^2}{3}\left(k_Fa\right)^7\frac{a^6}{(2\pi)^9}\int_0^2 d\bar{q}dcos\theta_q d\phi_q\left(I_2I_0+I_1^2\right).
\end{equation}

Evaluating $I_{2,1,0}$ and performing the integrals gives
\begin{equation}
\mathrm{Re}\sigma(\omega)\approx U^2m^2a^6 (k_Fa)^7 \frac{1}{(2\pi)^{9}}{0.24\pi}.
\label{eq:sigmafinal}
\end{equation} 

If we do similar calculations for two dimension where the rotation matrix is given by
\bea
R(\theta)=\begin{bmatrix}
\cos(\theta) & -\sin(\theta) \\
\sin(\theta) &\cos(\theta)
\end{bmatrix},
\eea
then evaluating the convolution will give us exactly zero, which we have shown in our previous work \cite{Mu22}.

For the DMFT approach, we only keep the self energy diagrams as in Fig.~\ref{fig:DMFTlattice} (each diagram has a spin degeneracy of two). In the low frequency limit since the self energy is proportional to $\omega^2$, by doing the integral over internal frequency, which gives $\omega^3$, this cancels $\omega^3$ in the prefactor of Eq.~(\ref{eq:conductivitydmft}). The conductivity will approach a constant.
Using the local self energy computed in Eq.~(\ref{eq:ImsigmaDMFTlowf}), we obtain the conductivity per unit cell $\mathrm{Re}\sigma(\omega)= U^2m^2a^6 (k_Fa)^6 \frac{(4\pi)^4}{(2\pi)^{12}}\frac{2}{9}\pi.$

\begin{figure}[h!]
\centering
\includegraphics{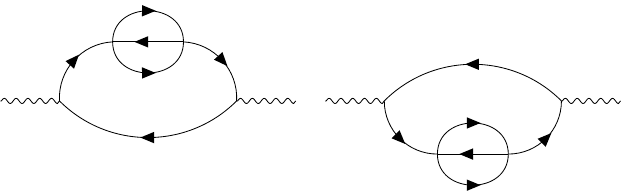}
\caption{Self energy contribution to the optical conductivity of the Hubbard model}
\label{fig:DMFTlattice}
\end{figure}

\section{Dirac system}
\label{app:3D Dirac}

\subsection{General remarks}
We decompose the non-interacting Green's function (a matrix in spin and orbital space) as
\bea
G_0(\vec{k},i\omega_n)&=&G_+(\vec{k},i\omega_n)+G_-(\vec{k},i\omega_n) \nonumber \\
&=&\frac{1}{2}\frac{1+\sigma_z\otimes \hat{k}\cdot\vec{\tau}}{i\omega_n+\mu-vk}+\frac{1}{2}\frac{1-\sigma_z\otimes \hat{k}\cdot\vec{\tau}}{i\omega_n+\mu+vk}.
\eea

Each diagram is a combination of Green function lines and interaction vertices; the form of the interaction we have chosen means that the sequence of lines corresponding to a definite spin corresponds to a  product of the orbital-space G matrices.

\subsection{Self energy}
This subsection gives some details of the perturbative self energy calculation for a Dirac system with a chemical potential near the Dirac point. Starting from the fundamental Eq.~(\ref{eq:Imsigma2}), specializing to $\omega,\mu>0$ and $T=0$, noting that the Green function lines labelled by $p_1$ and $p_2$ must lie in the upper band we have
\begin{equation}
\mathrm{Im}\Sigma^{\uparrow,ab}(\vec{k},\omega)=-\frac{U^2a^6}{(2\pi)^6}\frac{\pi}{8}\left(S^{+--}(k,\omega)+S^{+-+}(k,\omega)\right),
\label{eq:Imsigmadirac3}
\end{equation}
with
\begin{eqnarray}
\mathbf{S}^{+--}(k,\omega)&=&\int d\Omega_1 d\Omega_2 \mathbf{M}_+(\hat{p}_1)Tr\left[\mathbf{M}_-(\hat{p}_2)\mathbf{M}_-(\widehat{p_1+p_2-k})\right]\times
\label{eq:S+++}\\
&&\int_{k_F}^{\frac{\omega}{v}+k_F} p_2^2dp_2\int_{k_F}^{\frac{\omega}{v}+2k_F-p_2}p_1^2dp_1\delta\left(vp_1+vp_2-v\left|\vec{p_1}+\vec{p_2}-\vec{k}\right|-(\omega+\mu)\right),
\nonumber
\\
\mathbf{S}^{+-+}(k,\omega)&=&\int d\Omega_1 d\Omega_2 \mathbf{M}_+(\hat{p}_1)Tr\left[\mathbf{M}_-(\hat{p}_2)\mathbf{M}_+(\widehat{p_1+p_2-k})\right]\times
\label{eq:S++-}\\
&&\int_{k_F}^{\frac{\omega}{v}+k_F} p_2^2dp_2\int_{k_F}^{\frac{\omega}{v}+2k_F-p_2}p_1^2dp_1\delta\left(vp_1+vp_2+v\left|\vec{p_1}+\vec{p_2}-\vec{k}\right|-(\omega+\mu)\right),
\nonumber
\end{eqnarray}
and
\begin{equation}
    \mathbf{M}_\pm(k)=\mathds{1}\pm \hat{k}\cdot\vec{\tau}.
    \label{eq:Mdef}
\end{equation}
We observe that $S^{+-+}=0$ for $\omega<\mu$ but for $\omega\gg\mu$, $S^{+-+}\gg S^{+--}$.

Noting that $\omega+\mu=vk_\omega$ and  $k_\omega/k_F=1+\omega/\mu$, and scaling $p_1$ and $p_2$ by $k_\omega$ and defining $E_L=v/a$ we see that the imaginary part of the self energy can be explicitly written in the scaling form given in Eq.~(\ref{eq:Imsigma3a}).

For $\omega\ll\mu$, $S^{+-+}=0$ and we may set the magnitudes of $p_1$ and $p_2=k_F$; 
\bea
F_0(\frac{k}{k_F})&=&\int d\Omega_1d\Omega_2 (2+\hat{p_1}\cdot \hat{p_2}-\frac{\vec{k}}{k_F}\cdot \hat{p_2})\delta\left(1-\left|\hat{p_1}+\hat{p_2}-\frac{\vec{k}}{k_F}\right|\right), \label{eq:F0}\\
F_1(\frac{k}{k_F})&=&\int d\Omega_1d\Omega_2 (\hat{p_1}\cdot \hat{k})(2+\hat{p_1}\cdot \hat{p_2}-\frac{\vec{k}}{k_F}\cdot \hat{p_2})\delta\left(1-\left|\hat{p_1}+\hat{p_2}-\frac{\vec{k}}{k_F}\right|\right). \label{eq:F1}
\eea

For $\omega\gg\mu$ we may focus on $S^{+-+}$, set $\mu=0$ and obtain Eq.~(\ref{eq:IDirachigh}) with
\bea
G_0(\frac{vk}{\omega})&=&2\int p_1'^2dp_1'd\Omega_1p_2'^2dp_2'd\Omega_2 (1-\widehat{(p_1'+p_2'-k)}\cdot \hat{p_2})\nonumber \\
&\times&\delta\left(p_1'+p_2'-\frac{1}{\frac{vk}{\omega}}+\left|\vec{p_1'}+\vec{p_2'}-\hat{k}\right|\right), \label{eq:G0}\\
G_1(\frac{vk}{\omega})&=&2\int p_1'^2dp_1'd\Omega_1p_2'^2dp_2'd\Omega_2 (\hat{p_1'}\cdot \hat{k})  (1-\widehat{(p_1'+p_2'-k)}\cdot \hat{p_2})\nonumber \\
&\times&\delta\left(p_1'+p_2'-\frac{1}{\frac{vk}{\omega}}+\left|\vec{p_1'}+\vec{p_2'}-\hat{k}\right|\right).\label{eq:G1}
\eea
where we define $\displaystyle{p_{1}'=\frac{p_1}{k},p_2'=\frac{p_2}{k}}$.

\subsection{Optical Conductivity}
We sketch the evaluation of the diagrams shown  in Fig.~\ref{fig:diagrams}. The evaluation of the self energy diagrams  (type 1, panel (a) of Fig.~\ref{fig:diagrams}) proceeds slightly differently from the evaluation of the others. It is useful to denote the spin explicitly.  

\subsubsection{Self energy diagrams Fig.~\ref{fig:diagrams}(a)} In these diagrams we are explicitly computing the lifetime of a hole in the filled (-) band. Explicitly writng the analytical formula for the left-hand diagram gives
\bea
s_1(i\Omega_n)&=&\frac{a^3}{(i\Omega_n)^3}\int \frac{d^3\vec{k}}{(2\pi)^3}T\sum_{i\omega_n}(-4v^4)Tr[((\vec{\tau} \times \vec{k})\cdot \hat{z})G_-^{\uparrow}(\vec{k},i\omega_n+i\Omega_n)\Sigma^\uparrow(\vec{k},i\omega_n+i\Omega_n)\nonumber \\
&&G_-^{\uparrow}(\vec{k},i\omega_n+i\Omega_n)((\vec{\tau} \times \vec{k})\cdot \hat{z})G_+^{\uparrow}(\vec{k},i\omega_n)].
\eea

We insert the spectral representation $\displaystyle{\Sigma^\uparrow(\vec{k},i\omega_n+i\Omega_n)=\int \frac{dx}{\pi}\frac{-\mathrm{Im}\Sigma^\uparrow(\vec{k},x)}{i\omega_n+i\Omega_n-x}}$,
and perform the Matsubara sums and perform the analytical continuation and also take the trace over orbital indices.

Since we are focusing on low frequency,we may put all momenta on the Fermi surface, so that  we only need $\mathrm{Im}\Sigma^\uparrow(k_F\hat{k},\omega)$, which is given by explicitly evaluating Eq.~(\ref{eq:S+++}) as 
\bea
\mathrm{Im}\Sigma^\uparrow(k_F\hat{k},\omega)=-(\frac{4}{3}+\frac{4}{5}\hat{k}\cdot\vec{\tau})\frac{\omega^2}{\mu}\frac{U^2}{(\frac{v}{a})^2}\frac{(k_Fa)^4}{(2\pi)^6}\pi^3.
\eea


We obtain
\bea
s_1(\Omega)\sim \frac{1}{\Omega^3}\int dx \int d^3\vec{k} \delta(vk-\mu) (f(x-\Omega)-f(x))x^2 \frac{1}{(vk)^2} Tr[\cdots].
\eea

By performing the angular integral over $\vec{k}$  and collecting all the prefactors we obtain $\displaystyle{s_1(\Omega)=\frac{1}{(2\pi)^9}\frac{1}{2^6}\frac{4096}{135}\pi^4\frac{U^2}{(\frac{v}{a})^2}(k_Fa)^5 a^2}$.

\subsubsection{Type 2 vertex correction diagrams Fig.~\ref{fig:diagrams}(b)} In this diagram the dashed lines form a particle hole bubble which we denote by $B^{\downarrow}(\vec{q},i\nu)$ while the solid lines are a trace of products of G and the force operator, so that

\bea
s_2(i\Omega_n)&=&\frac{a^6}{(i\Omega_n)^3}\int \frac{d^3\vec{k}d^3\vec{q}}{(2\pi)^6}T\sum_{i\omega_n,i\nu}(4v^4)Tr[((\vec{\tau} \times \vec{k})\cdot \hat{z})G_-^\uparrow(\vec{k},i\omega_n+i\Omega_n)G_+^\uparrow(\vec{k}+\vec{q},i\omega_n+i\Omega_n+i\nu) \nonumber \\
&&((\vec{\tau} \times (\vec{k}+\vec{q}))\cdot \hat{z})G_-^\uparrow(\vec{k}+\vec{q},i\omega_n+i\nu)G_+^\uparrow(\vec{k},i\omega_n)]B^{\downarrow}(q,i\nu).
\eea

As with the self energy diagram we represent $B$ via a spectral representation, perform the analytical continuation and take the trace over the Green function indices obtaining 
\bea
s_2(\Omega)&\sim& \frac{1}{\Omega^3}\int dx\int d^3\vec{k}\int d^3\vec{q} (n(x)+f(vk-\mu+x+\Omega))(f(vk-\mu)-f(vk-\mu+\Omega)) \nonumber \\
&\times&\delta(v|\vec{k}+\vec{q}|-\mu) \mathrm{Im}B^\downarrow(q,x)\frac{1}{vk}\frac{1}{v|\vec{k}+\vec{q}|}Tr[\cdots].
\eea

In the low frequency limit both $\vec{k}$ and $\vec{k}+\vec{q}$ must be on the Fermi surface and 
\bea
\mathrm{Im}B^{\downarrow}(\vec{q},\nu)=-\frac{a^3}{(2\pi)^3}\frac{(2\pi)^2}{4}\frac{k_F^2}{v^2}\frac{\nu}{q}(2-\frac{q^2}{2k_F^2}),
\eea
where the range of $q$ is between $0$ and $2k_F$.

By doing the integral and the trace we obtain $\displaystyle{s_2=-\frac{3}{7}s_1}$.


\subsubsection{Type 3 and 4 vertex correction diagrams Fig.~\ref{fig:diagrams}(c,d)} In these diagrams the orbital structure is different; for example  the analytical expression for the left-hand type 3 diagram is
\begin{equation}
     s_3(i\Omega_n)=\frac{a^3}{(i\Omega_n)^3}\int\frac{d^3\vec{q}}{(2\pi)^3}T\sum_{i\nu}(4v)^4T_1(\vec{q},i\nu,i\Omega_n)T_2(\vec{q},i\nu,i\Omega_n),
     \label{eq:s31}
\end{equation}
with
\begin{eqnarray}
 T_1(\vec{q},i\nu,i\Omega_n)&=&T\sum_{i\omega_n}a^3\int\frac{d^3\vec{k}}{(2\pi)^3}Tr\left[G_+^{\uparrow}(\vec{k},i\omega_n)\left((\vec{\tau} \times \vec{k})\cdot \hat{z}\right)G_-^{\uparrow}(\vec{k},i\omega_n+i\Omega_n)G_+^{\uparrow}(\vec{k}+\vec{q},i\omega_n+i\nu)\right], \nonumber \\
    \label{eq:T1}\\
    T_2(\vec{q},i\nu,i\Omega_n)&=&T\sum_{i\omega_n}a^3\int\frac{d^3\vec{k}}{(2\pi)^3} Tr\left[G_+^{\downarrow}(\vec{k},i\omega_n+i\Omega_n)\left((\vec{\tau} \times \vec{k})\cdot \hat{z}\right)G_-^{\downarrow}(\vec{k},i\omega_n)G_+^{\downarrow}(\vec{k}+\vec{q},i\omega_n+i\nu)\right]. \nonumber \\
     \label{eq:T2}
\end{eqnarray}

Again we may take the trace, perform the frequency sums, analytically continue and evaluate the result in the low frequency limit where all the momenta must be near the Fermi surface.

\end{widetext}

\bibliographystyle{apsrev4-1}
\bibliography{DMFTlowdensityrefs.bib}

\end{document}